\documentclass[aps,noshowpacs,twocolumn,superscriptaddress]{revtex4-1}  
\usepackage{graphicx}  
\usepackage{dcolumn}   
\usepackage{bm}        
\usepackage{amssymb}   
\usepackage{amsmath}
\usepackage{amsfonts}
\usepackage{bibunits}
\usepackage{braket}
\usepackage{gensymb}
\usepackage{siunitx}

\defaultbibliography{flatten_aip_WSe2}
\defaultbibliographystyle{ieeetr}
\usepackage{setspace}

\begin{document}


\title{Microcavity enhanced single photon emission from two-dimensional WSe$_2$} 



\author{L.C.~Flatten} \email{lucas.flatten@materials.ox.ac.uk} \affiliation{Department of Materials, University of Oxford, Parks Road, Oxford OX1 3PH, United Kingdom} 
\author{L.~Weng} \affiliation{Department of Materials, University of Oxford, Parks Road, Oxford OX1 3PH, United Kingdom} 
\author{A.~Branny} \affiliation{Institute of Photonics and Quantum Sciences, SUPA, Heriot-Watt University, Edinburgh EH14 4AS, United Kingdom}
\author{S.~Johnson} \affiliation{Department of Materials, University of Oxford, Parks Road, Oxford OX1 3PH, United Kingdom}
\author{P.R.~Dolan} \affiliation{Department of Materials, University of Oxford, Parks Road, Oxford OX1 3PH, United Kingdom}
\author{A.A.P.~Trichet} \affiliation{Department of Materials, University of Oxford, Parks Road, Oxford OX1 3PH, United Kingdom}
\author{B.D.~Gerardot} \affiliation{Institute of Photonics and Quantum Sciences, SUPA, Heriot-Watt University, Edinburgh EH14 4AS, United Kingdom}
\author{J.M.~Smith} \email{jason.smith@materials.ox.ac.uk}  \affiliation{Department of Materials, University of Oxford, Parks Road, Oxford OX1 3PH, United Kingdom}


\date{\today}

\begin{bibunit}

\begin{abstract}
Atomically flat semiconducting materials such as monolayer WSe$_2$ hold great promise for novel optoelectronic devices. Recently, quantum light emission has been observed from bound excitons in exfoliated WSe$_2$. As part of developing optoelectronic devices, the control of the radiative properties of such emitters is an important step. Here we report the coupling of a bound exciton in WSe$_2$ to open microcavities. We use a range of radii of curvature in the plano-concave cavity geometry with mode volumes in the $\lambda^3$ regime, giving Purcell factors of up to 8 while increasing the photon flux five-fold. Additionally we determine the quantum efficiency of the single photon emitter to be $\eta = 0.46 \pm 0.03$. Our findings pave the way to cavity-enhanced monolayer based single photon sources for a wide range of applications in nanophotonics and quantum information technologies. 
\end{abstract}


\maketitle 


Single photon emission has been observed from a range of systems such as single atoms, quantum dots and localised excitons in a multitude of materials. Recently, two-dimensional semiconductors have attracted increased attention because of their strong interaction with light owed to a direct bandgap transition with a strong transition dipole moment of the delocalised exciton \cite{zhao_evolution_2013, xia_two-dimensional_2014}. Of these, the transition metal dichalcogenide WSe$_2$ has been found to contain localised excitons, stable at cryogenic temperatures below \SI{15}{\kelvin} \cite{he_single_2015,koperski_single_2015,srivastava_optically_2015,chakraborty_voltage-controlled_2015,tonndorf_single-photon_2015,kumar_strain-induced_2015}, emitting quantum light with impressive brightness \cite{kumar_resonant_2016} and stability \cite{iff_substrate_2017}. In particular the localised excitons can be created with nanometric precision \cite{kern_single-photon_2016,branny_deterministic_2017,palacios-berraquero_large-scale_2017}, exhibit strain tunability \cite{kumar_strain-induced_2015} and the hosting two-dimensional material allows for integration into ultra-compact, charge tunable devices \cite{xia_two-dimensional_2014, allain_electrical_2015}. \newline \indent

Optical microcavities enhance the light-matter interaction by concentrating the electric field \cite{vahala_optical_2003}. By reducing the number of states accessible to emitters coupled to the cavity and focusing the radiation into well defined modes they serve to improve the properties of single photon sources via the Purcell effect \cite{purcell_proceedings_1946,toishi_high-brightness_2009, kuhn_cavity-based_2010,buckley_engineered_2012}. As single photon sources are an essential component of quantum information technology, the engineering of suitable optical environments is therefore imperative \cite{beveratos_single_2002,obrien_photonic_2009}.  \newline \indent
In this article we describe the integration of a single photon emitter hosted by a monolayer of WSe$_2$ into a range of open-access cavities. In contrast to their monolithic counterparts such as micropillar structures, photonic crystals and ring resonators, open-access cavities allow in-situ tunability with respect to emitter position and wavelength by using two freely movable mirrors \cite{dolan_femtoliter_2010,
di_controlling_2012}. They thus enable integration of a versatile range of materials and both the coupling to the matter and the emission properties can be controlled with high precision \cite{dufferwiel_tunable_2015,
kaupp_purcell-enhanced_2016,
flatten_spectral_2016}. In particular, recent advances in the fabrication method with a Focused Ion Beam (FIB) allow small radii of curvature of the concave feature, thus enabling ultra-small mode volumes in the $\lambda^3$ regime \cite{dolan_femtoliter_2010,trichet_topographic_2015}. This in turn enables a sizable Purcell effect even in the bad emitter regime, where the homogeneous linewidth of the emitter is larger than the linewidth of the cavity mode \cite{di_controlling_2012,johnson_tunable_2015}.

\section{Monolayer WSe$_2$ in open microcavities}

The plano-concave microcavity consists of two opposing silica substrates, coated with a DBR comprising 13 pairs of $\text{SiO}_2/\text{TiO}_2$ with a reflectivity above $99.95\%$ at the center wavelength of $\lambda =$~\SI{740}{\nano\meter}. The mechanically exfoliated WSe$_2$ monolayer is transferred onto the planar mirror with a dry stamping process \cite{castellanos-gomez_deterministic_2014}. While this process leads to mostly multilayered WSe$_2$ stacks on the mirror surface, about five few micrometer sized spots showing monolayer properties were obtained on the sample (see Fig. \ref{fig1}a,b). These monolayers showed PL from both delocalised neutral excitons and bright, localised and spectrally narrow emitters visible below \SI{15}{\kelvin} (see PL map Fig. \ref{fig1}c). The bright spots could be identified as single photon emitters (SPEs) with emission wavelengths around \SI{750}{\nano\meter}, as described before \cite{he_single_2015,koperski_single_2015,srivastava_optically_2015,chakraborty_voltage-controlled_2015,tonndorf_single-photon_2015,kumar_strain-induced_2015,kumar_resonant_2016,iff_substrate_2017,kern_single-photon_2016,branny_deterministic_2017,palacios-berraquero_large-scale_2017}. To obtain a localised cavity mode (see Fig. \ref{fig1}d), another mirror with FIB-milled concave features with varying radii of curvature was positioned opposite of the SPE. Fig.~\ref{fig1}e shows cross sections of the concave features obtained with an AFM, demonstrating a range of radii of curvatures between \SI{3}{\micro\meter} and \SI{25}{\micro\meter} which were used in this study (the fabrication process has been published before  \cite{trichet_topographic_2015}). The cavity containing the exfoliated WSe$_2$ was integrated in a custom built confocal setup, enabling low temperature experiments with in-situ tunability of both mirrors relative to each other (see the supplementary material for more details on the setup).

Both mirrors forming the cavity were mounted in a tube within a liquid helium dewar, which provides optical access to the sample in our custom-build confocal microscope. The planar mirror was mounted facing downwards onto a large sample mount movable in three dimensions to vary the field of view and focus. Within this sample mount another stack of five piezo actuators (translation in three dimensions, tilt and rotation) provided the ability to align and position the featured mirror facing upwards relative to the planar mirror. 

\section{\label{results} Results}
\begin{figure}[h!]
\includegraphics[width=0.5\textwidth]{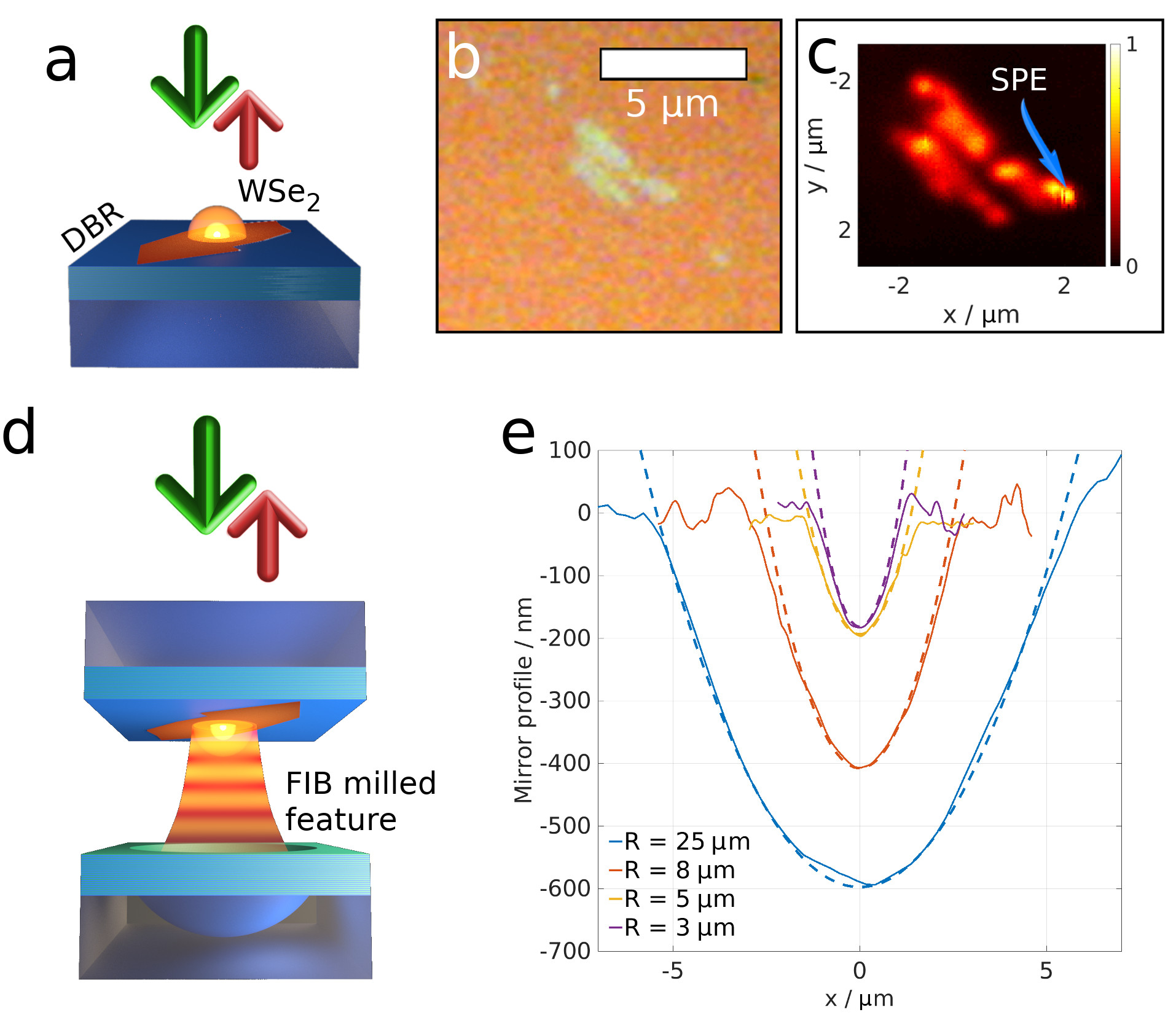}
\caption{Two-dimensional WSe$_2$ flakes in an open microcavity: a) A WSe$_2$ monolayer on a DBR mirror showing single photon emission from localised spots after excitation with a green laser. b) Optical microscope images of WSe$_2$ flakes on a mirror, deposited with a dry transfer method. c) Spectrally filtered PL signal (\SI{740}{\nano\meter}-\SI{760}{\nano\meter}) from the same flake as in \textbf{b} in a confocal microscope at \SI{4}{\kelvin}. The arrow points to a spot which shows single photon emission with a resolution limited spot-size of \SI{600}{\nano\meter}. d) The open-access microcavity setup is obtained by flipping the sample and positioning a mirror with concave features opposite it. The sample can be excited below the stopband of the mirrors with a green laser and emission is collected after emission through the flat mirror. e) AFM cross-sections of concave features forming one side of the microcavities, showing a range of radii of curvature. \label{fig1}}
\end{figure}

\begin{figure*}
\includegraphics[width=1\textwidth]{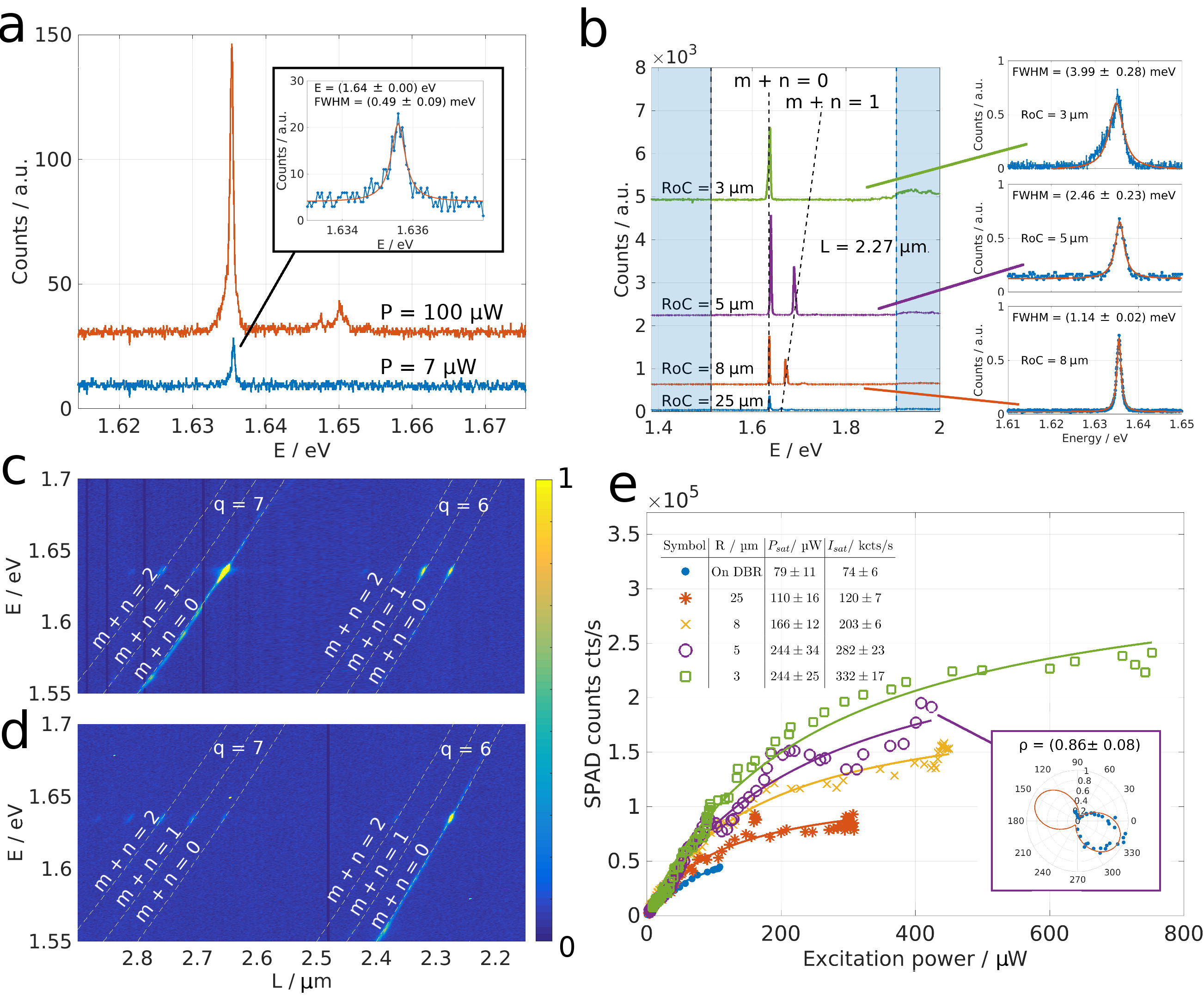}
\caption{Spectral properties of localised exciton outside and inside the cavity: a) Emission from spot marked in Fig. \ref{fig1}c for two different excitation powers $P$ with $\lambda_{\rm{exc}} = $ \SI{532}{\nano\meter}. The inset shows the lineshape of the emission spectrum at the lower power. b) Spectra of the cavity system for four different radii of curvature of the plano-concave resonator. For the concave feature with R $ = 25$,  \SI{8}{\micro\meter} (R $ = 5$,  \SI{3}{\micro\meter}) the TEM$^6_{0,0}$ (TEM$^5_{0,0}$) mode is brought into resonance with the SPE and the system is irradiated above saturation such that higher transverse ($m+n>0$) modes of the same longitudinal mode family are visible. The insets to the right show the lineshape of the fundamental mode for three cavities with a radius of curvature of \SI{3}{\micro\meter}, \SI{5}{\micro\meter} and \SI{8}{\micro\meter} from top to bottom. The shaded regions to the left and right of the main plot mark the end of the stop band of the mirrors. c) PL spectra of the cavity with R $ = $ \SI{8}{\micro\meter} as a function of the cavity length. The cavity is misaligned with respect to the SPE and thus the TEM$^6_{m+n=1}$ mode is populated. d) Same as in \textbf{c} with an aligned cavity. Now the TEM$^6_{0,0}$ is primarily populated and the TEM$^6_{m+n=1}$ remains dark. e) The saturation curves of the SPE in free space and aligned to the fundamental mode of different cavities. The table to the upper left summarises the parameters obtained from the saturation fits. The bottom inset shows the linearly polarised nature of the emission from the RoC $ = $ \SI{5}{\micro\meter} cavity.\label{fig2}}
\end{figure*}

Fig. \ref{fig1}a displays the sample facing upwards a custom built confocal setup, allowing for off-resonant excitation and collection of the PL. The flake was first localised within the optical microscope mode, obtained by illuminating the sample with a white LED from below and using a webcam to detect the transmitted photons as shown in Fig. \ref{fig1}b. Fig. \ref{fig1}c shows a PL map of the same region as shown in the optical microscope images, containing broad emission from most of the flake and a localised spot belonging to a single photon emitter with a FWHM of $\sim$~\SI{600}{\nano\meter} (marked with an arrow in Fig. \ref{fig1}c). For the PL map, the sample was excited with a continuous wave laser with $\lambda_{\rm{exc}} =$ \SI{532}{\nano\meter} and only photons between \SI{740}{\nano\meter} and \SI{760}{\nano\meter} were collected. The particular single photon emitter (SPE) which we study here has a central energy of $E_{\rm{SPE}} =$ \SI{1.64}{\electronvolt} and linewidth of $\Gamma_{\rm{SPE}}~\approx $~\SI{500}{\micro\electronvolt} (see Fig. \ref{fig2}a). While its spectral position has been stable over weeks, its brightness displayed fluctuations which increased with increased excitation power. The multilayered flakes visible in the optical microscope image in Fig. \ref{fig1}b and the PL from the rest of the WSe$_2$ flake allowed facile positioning of the SPE opposite of the concave FIB milled features and thus coupling to the cavity mode. Fig. \ref{fig2}b displays spectra from four different cavities for the TEM$_{0,0} $ mode brought into resonance with the SPE. To ascertain the correct cavity length and transverse position relative to the SPE, the cavity length was varied with one of the piezo microactuators and the PL spectrum (without any filtering) was collected. Fig. \ref{fig2}c,d shows two such cavity length scans, demonstrating bad transverse alignment (Fig. \ref{fig2}c) and optimal transverse alignment (Fig. \ref{fig2}d) to the SPE. Two successive mode families of the $R = $\SI{8}{\micro\meter} cavity with longitudinal index $q = 7,6$ are traversed with decreasing cavity length from left to right. The absolute cavity length can be obtained from the free spectral range and the coupling of the different transverse modes to the SPE are indicative of the quality of the lateral alignment. Since the electric field of the first transverse mode ($m+n = 1$) has a node in the center of its transverse electric field distribution, the optimal coupling to the SPE evidences itself with a minimal PL intensity for the first transverse mode and maximal intensity of the $m+n = 0$ mode. Note that the axes of the piezo microactuators show some crosstalk leading to varying lateral coupling quality for successive longitudinal modes. The different radii of curvature of the concave mirror lead to a different confinement of higher excited states. Thus the transverse mode spacing increases with decreasing radius of curvature, which is visible in the spectra plotted in Fig. \ref{fig2}b for four different cavities with $R = 25, 8, 5 $ and \SI{3}{\micro\meter}. To make these states visible in the PL spectrum in Fig. \ref{fig2}b the excitation power was increased well above saturation to $\approx $~\SI{1}{\milli\watt}. The three subplots in Fig. \ref{fig2}b show a more detailed spectrum of the TEM$_{0,0} $ mode for the three smallest cavities. In particular the lineshape of the smallest RoC cavity shows a substantial deviation from a Lorentzian lineshape, originating from diffraction losses at the edge of the feature and deviations of the profile. Additionally the spectrum of the $R =$ \SI{3}{\micro\meter} cavity only exhibits the TEM$^5_{0,0} $ mode, which we ascertained by scanning the cavity length over the next longitudinal mode family ($q=6$). This is indicative of a very shallow feature which does not allow confined transverse modes (the depth is about \SI{150}{\nano\meter} as shown in Fig. \ref{fig1}e). The SPE shows typical saturation behaviour as the excitation power is increased (see Fig. \ref{fig2}e). Care is taken to collect only emission from the TEM$_{0,0} $ mode of the different cavities. A first sign of a modulation of the electro-optical properties of the SPE due to the presence of the cavity, the Purcell effect, is the change in saturation counts in the four different photonic environments, which increases roughly five-fold from $I_{\rm{sat}} = (74\pm6)$ kcts/s at $P_{\rm{sat}} =$ \SI{79}{\micro\watt} to $I_{\rm{sat}} = (332\pm17)$ kcts/s at $P_{\rm{sat}} =$ \SI{244}{\micro\watt} (see table in inset in Fig. \ref{fig2}e). The emission is linearly polarised with a degree of polarisation of $p = (0.86 \pm 0.08)$, shown in the bottom inset for the cavity with a RoC of \SI{4}{\micro\meter}.

\begin{figure*}
\includegraphics[width=1\textwidth]{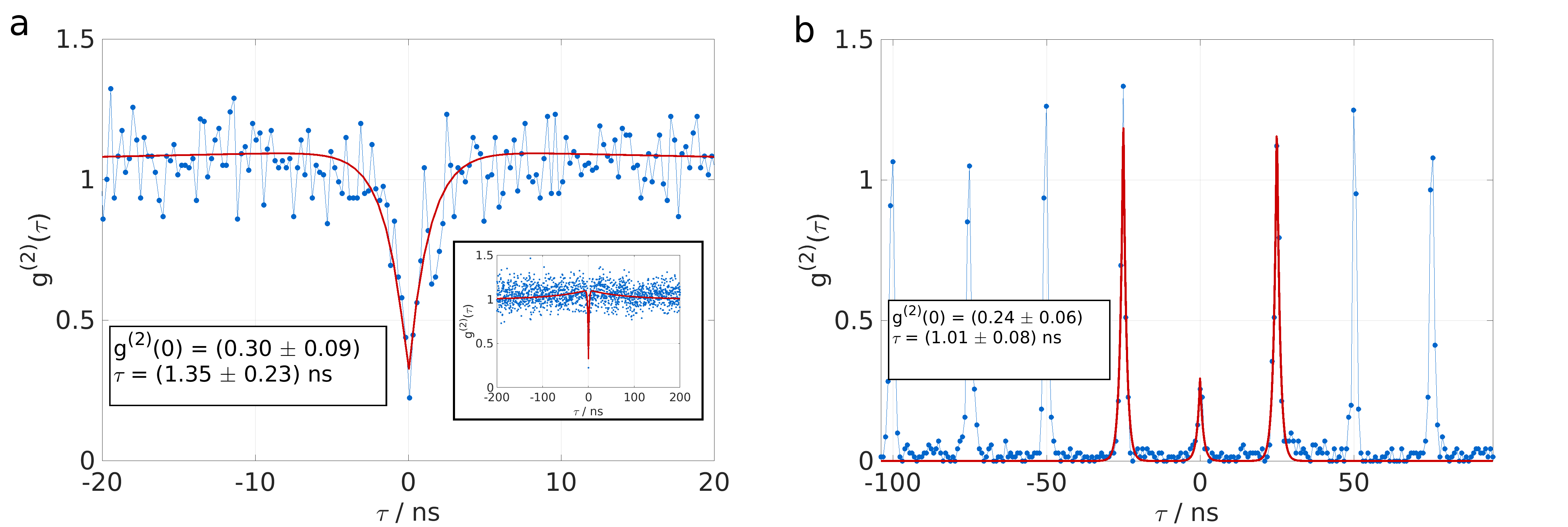}
\caption{Second order correlation function obtained from a Hanbury-Brown and Twiss setup: The emission from the TEM$^6_{0,0}$ mode of the RoC $ = $ \SI{3}{\micro\meter} cavity is collected with a) continuous wave excitation and b) pulsed excitation with $f_{\rm{rep}} = $ \SI{40}{\mega\hertz} at saturation power. The inset in a) shows the same data on a larger timescale, demonstrating a weak signature of photon bunching.\label{fig3}}
\end{figure*}

Additional to the typical saturation curve, the photon statistics of the emission should show anti-bunching as the evidence for single photon emission. To prove this, Fig. \ref{fig3}a,b show the second order autocorrelation function $g^{(2)}(\tau)$ for both continuous wave (a) and pulsed excitation (b) with $f_{\rm{rep}} = $ \SI{40}{\mega\hertz} at $\lambda_{\rm{exc}} = $ \SI{532}{\nano\meter} below saturation. We have aquired data from all cavities and consistently obtained $g^{(2)}(0)$ values below 0.5 without background subtraction. The data presented in Fig. \ref{fig3} stems from emission from the smallest cavity with an RoC of \SI{3}{\micro\meter} without background subtraction. We have observed signs of bunching on larger timescales of approx. $\SI{75}{\nano\second}$ (see inset in Fig. \ref{fig3}a, the experimental parameters are given in the supplementary material). The weak bunching characteristics could be indicative of a long-lived shelving state, however an alternative origin was identified previously in the spectral jittering of the emitting state \cite{koperski_single_2015}. For the pulsed measurement, the $g^{(2)}(0)$ value is obtained as the ratio of the areas of the central peak to the other peaks. Note that the time constant in both autocorrelation measurements is roughly $\tau = $ \SI{1}{\nano\second}, shown later to be a result of the Purcell enhancement of the small cavity.

\begin{figure*}
\includegraphics[width=1\textwidth]{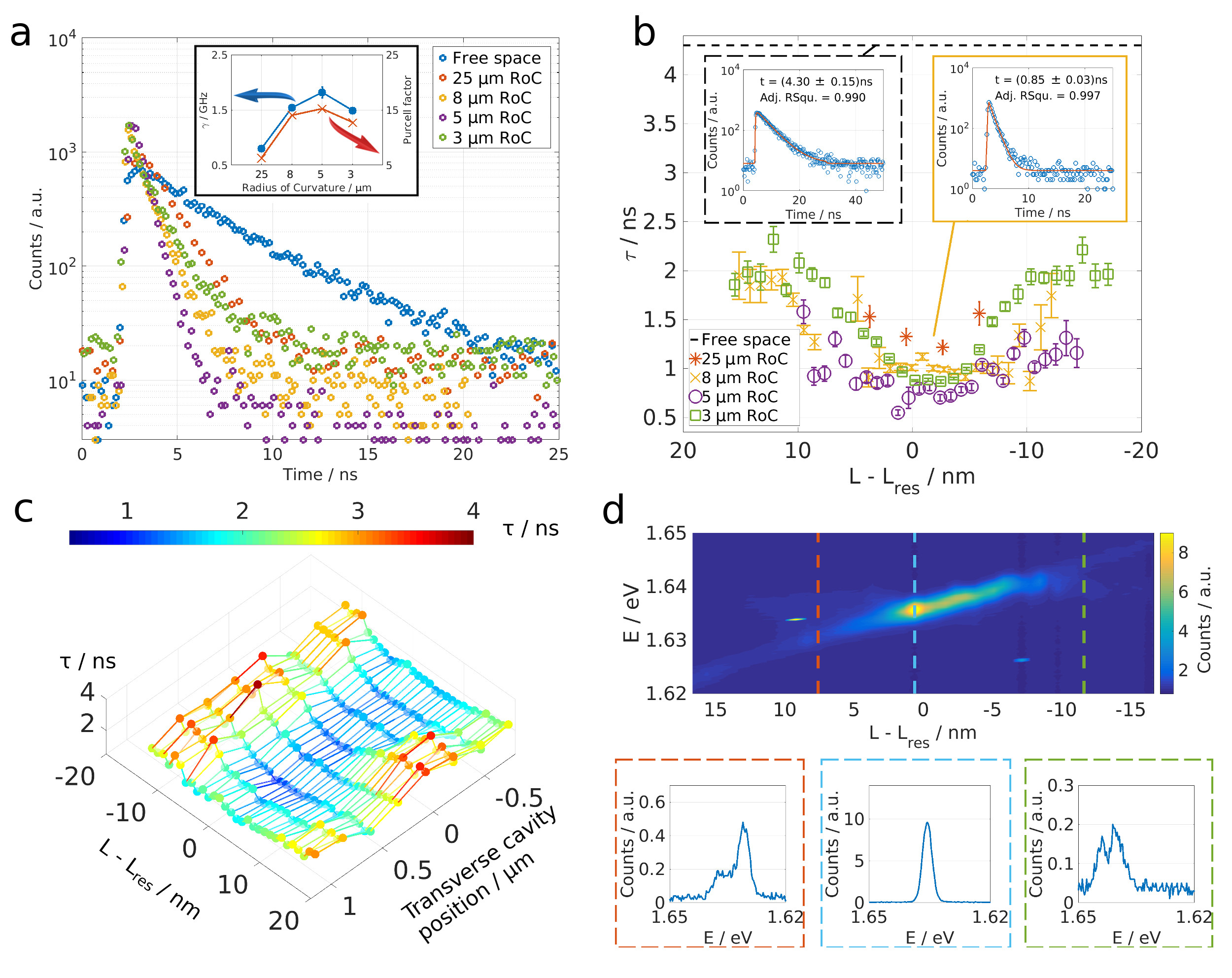}
\caption{Control over the Purcell effect in the open cavity: a) Time-resolved photoluminescence from SPE in four different cavities and without confining mirror showing Purcell enhancement of the radiative decay. Inset: Largest radiative decay constant measured for the respective radius of curvature and theoretical Purcell enhancement. b) The lifetime as the cavity length is varied such that the TEM$_{0,0}$ traverses the SPE for four different cavities. The reference lifetime $\tau_{\rm{hs}}$, where the SPE emits into the half space above the mirror, is obtained by the fit in the left inset and marked with the dashed black line in the main plot. The right inset shows the fit to the time resolved PL signal from the SPE in the $R = $ \SI{5}{\micro\meter} cavity on resonance. c) Effect of spatial detuning: Lifetime of SPE in $R = $ \SI{3}{\micro\meter} cavity as a function of both the cavity length and the cavity's transverse position relative to the SPE. d) Top panel: PL spectra from SPE in the $R = $ \SI{5}{\micro\meter} cavity as the cavity length is varied. The horizontal position coincides with the data presented in Fig. \ref{fig4}b. Lower panel: Three selected spectra from data presented above, corresponding to the vertical, dashed lines in the top plane plot. Note the scale difference when comparing the middle plot on resonance with the two detuned cases (differing by a factor of 20 and 50 respectively).  \label{fig4}}
\end{figure*}

The radiative decay rate of the excited state of the SPE can be measured more directly by detecting the time between excitation pulse and photon emission in a time resolved photoluminescence (TRPL) experiment. Fig. \ref{fig4}a shows this data for different photonic environments. The dataset with the longest lifetime $\tau$ (slowest radiative decay rate $\gamma = \frac{1}{\tau}$) was taken for the situation depicted in Fig. \ref{fig1}a with the SPE located on the DBR emitting into the half-space above. The data reveals a lifetime of $\tau_{\rm{hs}} = (4.3 \pm 0.2)$ \SI{}{\nano\second} in this case. Bringing another mirror into a position opposite of it forms the cavity and alters the density of optical states, into which the emitter can irradiate. The other datasets in Fig. \ref{fig4}a show the TRPL signal from the SPE tuned into resonance with the TEM$_{0,0}$ modes of the four different cavities. The numeric values of the decay rate are plotted in the inset as a function of the radius of curvature of the concave feature (left ordinate) and compared with the theoretical Purcell enhancement given by $F_{\rm{theo}} =  3 \lambda ^3 \mathcal{Q}_{\rm{eff}} / 4 \pi ^2 V $ (right ordinate), where $\mathcal{Q}_{\rm{eff}}$ is the effective quality factor and $V$ is the mode volume (more information are given in the supplementary material). Fig. \ref{fig4}b presents additional datapoints, obtained by descreasing the cavity length from left to right, such that the TEM$^6_{0,0}$ (TEM$^5_{0,0}$) energetically traverses the SPE energy for concave features with R $ = 25$,  \SI{8}{\micro\meter} (R $ = 5$,  \SI{3}{\micro\meter}). The two insets show fits to the TRPL data, in which the instrument response function of the single photon detector has been convoluted with a monoexponential function, giving the time constant $\tau$. The left inset displays data from the SPE emitting into free space, the left one data from the SPE tuned into resonance with the TEM$^5_{0,0}$ mode of the $R = $ \SI{5}{\micro\meter} cavity. The data in the main plot is obtained by similar fits and shows a dip in lifetime (or a peak in the radiative decay rate) as the cavity lengths of the four different cavities are tuned, such that the TEM$_{0,0}$ mode is spectrally scanned across the SPE emission. While the presence of all cavities has a marked influence on the lifetime, the largest RoC cavity results in a minimal lifetime of $\tau = (1.22 \pm 0.07)$ \SI{}{\nano\second} and the best enhancement is seen for the $R = $ \SI{5}{\micro\meter} cavity, where the lifetime is $\tau = (0.55 \pm 0.03)$ \SI{}{\nano\second}. Fig. \ref{fig4}d displays a series of spectra for the same cavity length changes as in Fig. \ref{fig4}b for the $R = $ \SI{5}{\micro\meter} cavity. As the separation between the mirrors is decreased the energy of the cavity mode increases and traverses the SPE, with the resonant case depicted in the middle inset in the lower panel (light blue, dashed line). The left (right) subplot of the lower panel marked with a red dashed line (green dashed line) show spectra, where the cavity mode is detuned from the emitter to lower (higher) energies. The peak structure for the detuned cases results from the cavity environment, which enhances off-peak parts of the underlying delocalized exciton PL resonant with the mode and suppresses the SPE emission, for whose energy no cavity state is available.

Along with the spectral alignment of the emitter to the cavity mode, the position of the emitter in the electric field of the mode is another critical parameter for the Purcell effect. While the placement of the monolayer at the surface of the planar DBR with a low refractive index termination ensures longitudinal alignment to an antinode of the mode, the transverse alignment is a parameter which can be optimised experimentally in situ (as demonstrated in Fig. \ref{fig2}c). Fig. \ref{fig4}c shows the effect that the spatial detuning has on the magnitude of the Purcell enhancement. We plot the lifetime as a function of both spectral (cavity length) and spatial detuning (along an arbitrary direction of the azimuthally symmetric concave cavity parallel to the mirror surface). While the lifetime values for little spatial detunings range from $\tau = 0.9$ \SI{}{\nano\second} to $\tau = 3.4$ \SI{}{\nano\second} the Purcell effect causes lower variations for large spatial detunings, where the lifetimes lay in the range of $\tau = 1.5$ \SI{}{\nano\second} to $\tau = 3.0$ \SI{}{\nano\second}.

\setlength{\tabcolsep}{8pt}
\begin{table*}
\centering
\begin{tabular}{c|c|c|c|c|c|c|c}
$R_{\rm{phys}}$ / \SI{}{\micro\meter} & $R_{\rm{opt}}$ / \SI{}{\micro\meter} & $\Gamma_{\rm{cav}}$ / meV & $\mathcal{Q}_{\rm{eff}}$ & $V / \lambda^3$ & $F'_{\rm{theo}}$ & $f$ & $\eta$ \\ \hline
25 & 26.5 & 2.15 & 607 & 7.35 & 7.3 & 3.44 & $0.43$\\
8 & 14.7 & 1.14 & 970 & 5.26 & 15.0 & 6.62 & $0.44$\\
5 & 7.6 & 2.46 & 545 & 2.72 & 16.2 & 7.81 & $0.50$\\
3 & (5.3) & 3.99 & 361 & (2.10) & (14.1) & 6.4 & $0.46$
\end{tabular}
\caption{Purcell-effect related quantities for the SPE coupled to the TEM$^6_{0,0}$ (TEM$^5_{0,0}$) mode of four cavities with R $ = 25$,  \SI{8}{\micro\meter} (R $ = 5$,  \SI{3}{\micro\meter}) at $L \approx$~ \SI{2.3}{\micro\meter} ($L \approx$~ \SI{1.9}{\micro\meter} ). R$_{\rm{phys}}$ and R$_{\rm{opt}}$ are the radii of curvature obtained by AFM measurement (comp. Fig. \ref{fig1}e) and as inferred from the transverse mode spacing in the analytic Gaussian beam description. $\Gamma_{\rm{cav}}$, $\mathcal{Q}_{\rm{eff}}$ and $V$ are the mode linewidth, effective quality factor of the cavity-emitter system and mode volume, which result in the effective theoretical value for $F'_{\rm{theo}} = \frac{3 \lambda ^3 \mathcal{Q}_{\rm{eff}}}{4 \pi ^2 V} + 1$. $f$ denotes the measured ratio of the excited state lifetime when emitting into the half-space above the dielectric mirror and within the cavity $f = \frac{\tau_{\rm{hs}}}{\tau_{\rm{cav}}}$. $F'_{\rm{theo}}$ and $f$ can be used to calculate the quantum efficiency $\eta$ of the SPE, which is shown in the last column (we derive the relation in the supplementary material). The values in brackets in the bottom row are inferred by assuming $\eta = 0.46$ for the emitter in the smallest cavity, as no transverse modes were visible to calculate $R_{\rm{opt}}$.}
\label{tab1}
\vspace{10pt}
\end{table*}

Tab. \ref{tab1} summarises the most important quantities for the four different cavities. The first column shows $R_{\rm{phys}}$, the physical radius of curvature as measured with an AFM (see Fig. \ref{fig1}e). $R_{\rm{opt}}$, the parameter in the second column, is the radius of curvature obtained from the transverse mode spacing (the analytical description can be found in the supplementary material). It describes the effective confining potential and has been reported before to exceed the nominal RoC for small curvatures \cite{trichet_topographic_2015}. $\Gamma_{\rm{cav}}$, $\mathcal{Q}_{\rm{eff}}$ and $V$ are the cavity mode FWHM linewidth, quality factor and volume that we infer from the optical radius of curvature. With these values we calculate the effective Purcell factor $F'_{\rm{theo}} = F_{\rm{theo}} +1 = \frac{3 \lambda ^3 \mathcal{Q}_{\rm{eff}}}{4 \pi ^2 V} + 1$ and compare it to the experimental value $F'_{\rm{exp}} = \frac{ \tau_{\rm{fs}}}{\tau_{\rm{cav}}}$. In  the experiment, $F'_{\rm{exp}}$ is not directly accessible because the emitter is deposited on the surface of the DBR, which influences the optical density of states even without an opposing mirror \cite{barnes_fluorescence_1998}. The excited state lifetime of $\tau_{\rm{hs}} = (4.3 \pm 0.2)$ is therefore not the true free space lifetime. In fact, the relationship between the measured ratio of the lifetimes $f = \frac{\tau_{\rm{hs}}}{\tau_{\rm{cav}}}$ and $F'_{\rm{exp}}$ is given by $F'_{\rm{exp}} = f(\mathcal{E}-1)\eta +f$ (we derive the given relations in the supplementary material). Here $\mathcal{E}$ is a factor that describes the enhancement (or suppression) of the decay rate caused by the DBR (for the low refractive index terminated mirror used in the experiment we find $\mathcal{E} \approx 1.2$ through finite-difference time-domain (FDTD) simulations \cite{OskooiRo10}).

\section{\label{discussion} Discussion}
Our experiments allow us to infer the quantum efficiency of the emitter and to estimate the losses in our cavity setup. The reported changes in the excited state lifetime of the three largest cavities together with the respective theoretical Purcell factor result in similar quantum efficiencies of the emitter with an average of $\eta = 0.46 \pm 0.03$. 

The small spread of the individual values for $\eta$ validates the assumption that the effective RoC $R_{\rm{opt}}$ and the associated mode volume describe the experimental situation. Would we have calculated the Purcell factor from the physical RoC, the values for $\eta$ would range between $ 0.25$ and $ 0.42$ with an average of $\eta_{\rm{phys}} = (0.35 \pm 0.06)$. For the smallest feature we could not establish a transverse mode spacing and thus could not infer $R_{\rm{opt}}$. Working backwards however and assuming $\eta = 0.46$ we obtain $R_{\rm{opt}} = 5.3 $ \SI{}{\micro\meter} in good agreement with previous reports \cite{trichet_topographic_2015}. The origin of the reduced quantum efficiency is unclear and could be elucidated by further experiments with emitters exhibitting smaller linewidths.

The saturation excitation power $P_{\rm{sat}}$ and countrate $I_{\rm{sat}}$ depend on the radiative decay rate $\gamma_{\rm{r}}$ only and should thus show enhancements proportional to the Purcell factor. The data presented in the table in Fig. \ref{fig2}e supports this theory weakly and suggests severe losses from the cavity mode. The same lens with an $\rm{NA} = 0.82$ was used for both the cavity coupling and free-space experiments. In the free-space experiment with the SPE emitting into the halfspace above the mirror this results in a collection efficiency of $\eta_{\rm{fs}} = 42.8 \%$. The divergence of the cavity modes in this study is below $\theta_{\rm{div}} = 0.4$, which suggests that all the light is collected from the cavity mode irradiating through the planar mirror side. Since the reflectivity of this side of the cavity is lower than the opposing patterned side, we obtain a collection efficiency of $\eta_{\rm{cav}} = 62.5 \%$ in this case (we give further information in the supplementary material). Assuming $P_{\rm{sat}} \propto F'_{\rm{theo}}$ we find a ratio of $\frac{2 P_{\rm{sat}}}{P_{\rm{sat \; fs}} F'_{\rm{theo}} }= 44\%,\, 30\%,\, 41\%$ and $48 \%$ for the four cavities from largeest to smallest. For the $I_{\rm{sat}}$ values we calculate a similar ratio, but subtract it from 1 to obtain losses (ie. deviations from the theoretically enhanced free-space saturation counts) of $\delta_{\rm{DBR}}(758 \SI{}{\nano\meter}) = 1 - \frac{I_{\rm{sat}} \; \eta_{\rm{cav}}}{I_{\rm{sat \; fs} \; \eta_{\rm{fs}}} F'_{\rm{theo}} }= 62\%,\, 71\%,\, 63\%$ and $49 \%$ for the four cavities in the same order. The spread in these values confirms our theory of proportionality only weakly. Origins of the losses at the DBR interface could be an increased surface roughness and/or contamination from the backside, where the mirror was attached to the sample holder. Note that the cavity linewidth exceeds the emitter linewidth for all cavities. Even though we see such losses our cavity setup still increases the photon flux of the SPE by $450 \%$ and increases the frequency of single photon emission by a factor of $7.8 $ when compared to irradiation into free space.

\section{\label{conclusion} Conclusion}
In conclusion we have presented a single photon emitter hosted by the transition metal dichalcogenide WSe$_2$ coupled to a range of FIB milled microcavities with mode volumes in the $\lambda^3$ regime. We have shown an enhancement of the spontaneous emission rate by a factor of 7.8, corresponding to an effective Purcell factor of $16.2$ for a cavity with a radius of curvature of $R = $ \SI{5}{\micro\meter} together with full spectral and spatial tunability. Coupling to the cavity has increased the saturation countrate from 74~$\frac{\rm{kcts}}{\rm{s}}$ on the mirror surface to 332~$\frac{\rm{kcts}}{\rm{s}}$ within the cavity, an almost five-fold enhancement. 
The ability to integrate quantum emittters in 2D materials with open-access microcavities offers a promising route towards in-situ tunable, modular and ultra-bright single photon sources for a wide range of applications in quantum information technology.

\begin{acknowledgments}
L.C.F. acknowledges support from the European Commission (project WASPS, 618078). A.B. acknowledges support from EPSRC grant EP/L015110/1. B.D.G. acknowledges support from an ERC Starting Grant (no. 307392) and a Royal Society University Research Fellowship.
\end{acknowledgments}

\section*{Supplementary Material}
The supplementary material includes further details of the sample preparation and emitter properties, the analytical description of the cavity modes, expressions for the Purcell effect and the quantum and collection efficiency and the expression for the second-order degree of coherence; original data sets available at https://ora.ox.ac.uk/. The authors declare no competing financial interest.

\putbib

\end{bibunit}

\clearpage
\pagebreak
\widetext
\appendix
\begin{center}
\textbf{\large Supplementary materials: Microcavity enhanced single photon emission from two-dimensional WSe$_{\textrm{2}}$}
\end{center}
\setcounter{equation}{0}
\setcounter{figure}{0}
\setcounter{table}{0}
\setcounter{page}{1}
\makeatletter
\renewcommand{\theequation}{S\arabic{equation}}
\renewcommand{\thefigure}{S\arabic{figure}}
\renewcommand{\bibnumfmt}[1]{[S#1]}
\renewcommand{\citenumfont}[1]{S#1}
\begin{bibunit}
\section{Sample preparation}

The plano-concave microcavity consists of two opposing silica substrates, into one of which a concave feature has been milled with a Focused Ion Beam. The subtrates are coated with a DBR comprising 13 pairs of $\text{SiO}_2/\text{TiO}_2$, the planar mirror ending with a low refractive index $\text{SiO}_2$ layer. The featured mirror does not have the last layer and thus ends with the high refractive index material $\text{TiO}_2$. The planar (patterned) mirror thus has a reflectivity of $99.95\%$ ($99.97\%$) at the center wavelength of $\lambda =$~\SI{740}{\nano\meter}, corresponding to a maximum achievable finesse $\mathcal{F} \approx 5000$. For a cavity with mirror reflectivities $R_i$ and $1-R_i \ll 1$ the total transmission through side $i = 1,2$ can be approximated with $T_i^{\rm{tot}} = \frac{1-R_i}{1-R_1 R_2}$ (by summing the reflected and transmitted parts and assuming a resonant cavity mode). For the presented configuration one obtains $T^{\rm{tot}} = 62.5 \%$ through the lower reflectivity, planar cavity mirror.

The WSe$_2$ monolayer is created via mechanical exfoliation from bulk WSe$_2$ and transferred onto the planar DBR with a dry stamping process. Most of the deposited material has several layers of WSe$_2$, only a few monolayer flakes with extensions over a few micrometers could be obtained in one exfoliation process and were found via optical microscope inspection. 
\section{Cavity modes}
Cavity modes can be described with solutions to the paraxially approximated Helmholtz equation with the correct boundary conditions. In three dimensions, each mode can be classified by a set of three natural numbers $q$, $m$ and $n$, which are called longitudinal ($q$) and transverse ($m$ and $n$) mode numbers leading to the shorthand notation TEM$^q_{m,n}$ for such a mode. The analytical solution for the frequency of each mode is:
\begin{equation}
\nu^q_{m,n} = \frac{c}{2L}\left( q+\frac{1+m+n}{\pi} \arccos(\sqrt{g_1 g_2}) \right)
\label{eq3}
\end{equation} 
where $g_i = 1-L/R_i$ for both mirrors and $R_i$ the radius of curvature of the respective mirror. In our system we set $g_1 = 1$ and $g_2 = 1 - L/R$. The transverse mode spacing, from which we can infer the radius of curvature, is the wavelength difference of two successive modes with wavelength $\lambda_1$ and $\lambda_2$ and $\Delta m+\Delta n = 1$. Solving Eq. \ref{eq3} for R in this situation we obtain (with $\lambda_2 > \lambda_1$):
\begin{equation}
R = L \left[ \sin \left( \frac{2\pi L(\lambda_2 - \lambda_1)}{\lambda_1 \lambda_2}\right) \right]^{-2}
\label{eq4}
\end{equation}

\begin{figure}[h!]
\includegraphics[width=0.6\textwidth]{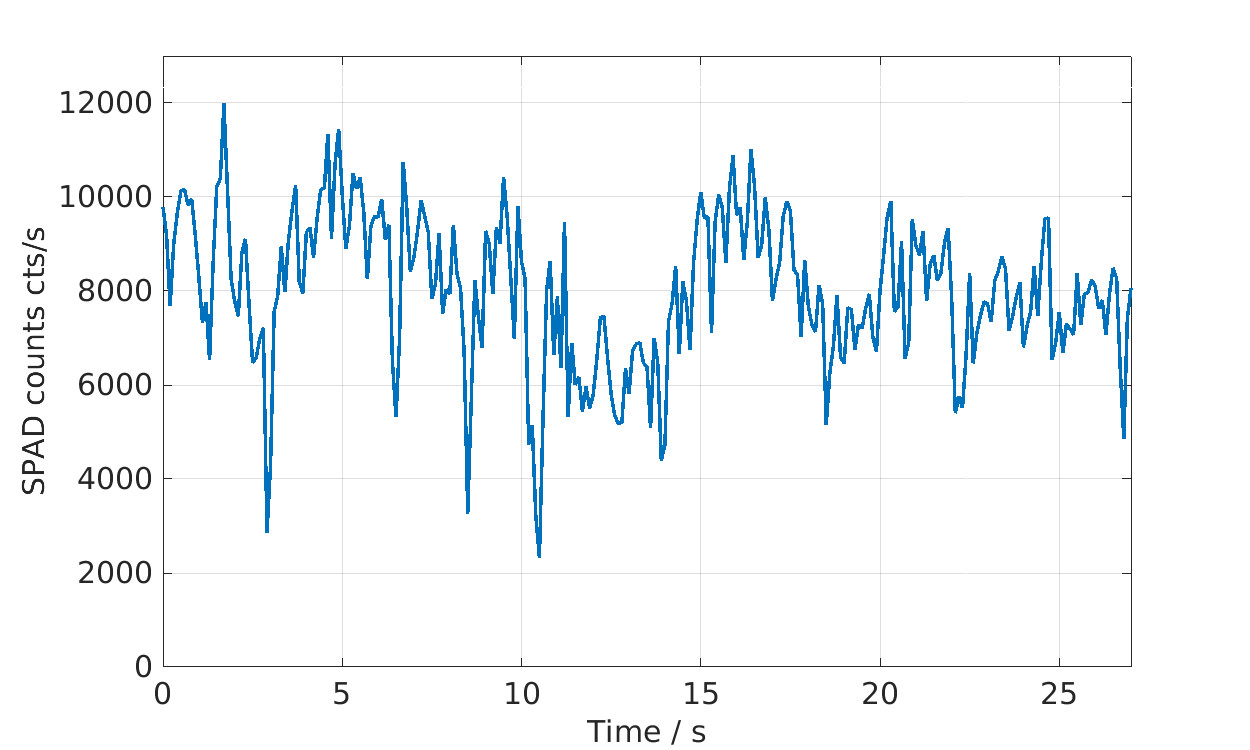}
\caption{Time trace of signal of WSe2 SPE in the RoC \SI{5}{\micro\meter} cavity. The photons are detected by a SPAD and the countrate is queried at a rate of \SI{10}{\hertz}. \label{figs2}}
\end{figure}

\section{Emitter brightness fluctuations}
The emitter brightness fluctuates with higher excitation powers, an effect frequently observed in single quantum dot systems where it is attributed to electric charge fluctuations in the vicinity of the emitter, caused by the non-resonant laser excitation. The fluctuation in charges then causes spectral shifts (due to modification of the confining potential and the Stark effect) and periods of non-radiative recombination (blinking, caused by Auger processes). For the emitters in WSe2 under study, previous publications reported increased spectral shifting with increased excitation powers (see for ex. Fig. 2c,d in \cite{koperski_single_2015}). When placed in a cavity with a mode linewidth below the spectral shifting range, intensity fluctuations are therefore to be expected. Fig. \ref{figs2} shows a time trace of the emitter signal obtained by consecutive SPAD measurements. It details that the average countrate over seconds is constant, but the subsecond variance is of order of 50$\%$ of the signal.

\section{Purcell effect}
In general, the spontaneous emission rate of an emitter is governed by the transition matrix element from initial to final state and the density of final states. A change from free space to a confined  environment causes a modulation of the density of optical states (the Purcell effect). In particular the coupling to a single cavity mode causes a change in spontaneous decay rate by the Purcell factor \cite{purcell_proceedings_1946, fox_quantum_2006}:
\begin{align}
f_{\rm{theo}} = \frac{\gamma_{\rm{cav}}}{\gamma_{\rm{fs}}} = \frac{\tau_{\rm{fs}}}{\tau_{\rm{cav}}} &= \frac{3 \lambda ^3 \mathcal{Q}_{\rm{eff}}}{4 \pi ^2 V} \Big\{ \left( \frac{\delta \omega^2}{4(\omega - \omega_0)^2+\delta \omega ^2} \right) \left(\frac{|\vec{p}\cdot \vec{E}|}{|\vec{p}||\vec{E}|}\right)^2 \left( \frac{|\vec{E} ( \vec{r})|}{|\vec{E}_{max}|}\right)^2 \Big\} \\
&= \quad F_{\rm{theo}} \; \Big\{ \upsilon  \xi  \chi \Big\}
\label{eqPurc}
\end{align}
Here $\tau_{\rm{fs}}$ and $\tau_{\rm{cav}}$ are the lifetimes in free space and within the cavity respectively. The two most important quantities in the expression are $\mathcal{Q}_{\rm{eff}}$, the quality factor of the system and $V$, the volume of the cavity mode. The expressions in the brackets quantify the spectral detuning $\upsilon$, the dipole alignment $\xi$ and the position of the emitter relative to the electric field $\chi$ respectively and should be close to unity in the optimised case. $F_{\rm{theo}} = \frac{3 \lambda ^3 \mathcal{Q}_{\rm{eff}}}{4 \pi ^2 V}$ thus gives the magnitude of the Purcell enhancement in this case. $\mathcal{Q}_{\rm{eff}}$ and $V$ are given by
\begin{equation}
\mathcal{Q}_{\rm{eff}} = \frac{\lambda}{\delta \lambda_{\text{cav}}+\delta \lambda_{\text{em}}} 
\qquad \& \qquad
V = \frac{\lambda L^2}{4}\sqrt{\frac{R}{L}-1}
\label{eq2}
\end{equation}
where $\delta \lambda_{\text{em}}$ and $\delta \lambda_{\text{cav}}$ are the linewidths of the emitter and the cavity respectively. $\lambda$ is the wavelength of the cavity mode, which on resonance is the same as the wavelength of the emitter. In the expression for $V$, $L$ is the cavity length and $R$ is the radius of curvature of the concave mirror. To account for the fact that the cavity is open and a coupling to free space modes is possible within good approximation, the effective Purcell factor $F'_{\rm{theo}} = F_{\rm{theo}} + 1 = \frac{3 \lambda ^3 \mathcal{Q}_{\rm{eff}}}{4 \pi ^2 V} + 1$ is introduced.

\section{Spontaneous emission in the vicinity of dielectric surfaces}
\begin{figure}
\includegraphics[width=0.6\textwidth]{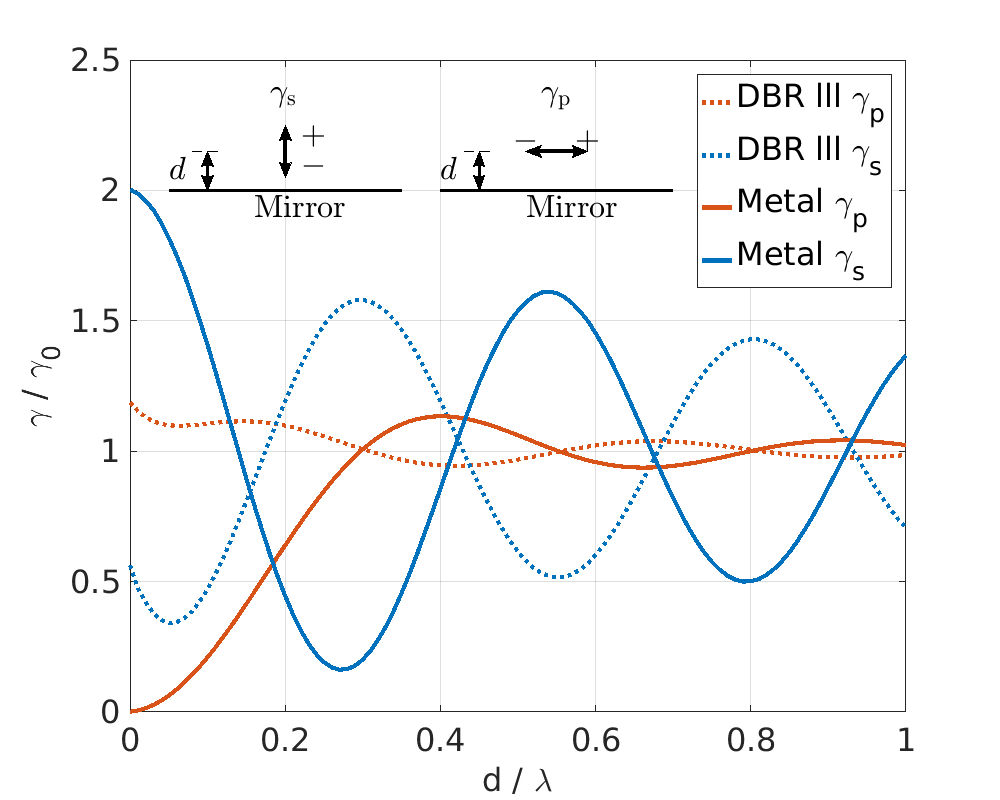}
\caption[flushleft]{Spontaneous emission rate of a dipole in front of a perfect dielectric mirror with a low refractive index last layer (DBR lll) and for a perfect metal mirror for different orientations of the dipole. As the distance $d$ between dipole and mirror is varied the emission rate shows an oscillatory behaviour. For separations exceeding multiple wavelengths of the emission the free space properties of the emitter are recovered. Simulations were performed with the finite-difference time-domain (FDTD) method, using a freely available software package \cite{OskooiRo10}. \label{figS3}}
\end{figure}

The Purcell effect is normally associated with a cavity-emitter systen. However, a similar enhancement or suppression in spontaneous emission rate can be observed for an emitter close to a dielectric interface originating from the change in the optical density of states \cite{barnes_fluorescence_1998}. For an emitter oriented perpendicular (s) or parallel (p) to the the surface of a dielectric mirror ending with a low refractive index layer, the emission will be modified as shown in Fig. \ref{figS3} (dotted lines). The continuous lines show the spontaneous emission rate of the emitter when a perfect metal mirror is used. All four datasets were obtained with finite-difference time-domain (FDTD) simulations, using a freely available software package \cite{OskooiRo10}.
\section{Quantum efficiency}
The quantum efficiency of an emitter is defined as $\eta = \frac{\gamma_{\rm{r}}}{\gamma_{\rm{r}} + \gamma_{\rm{nr}}}$, where $\gamma_{\rm{r}}$ ($\gamma_{\rm{nr}}$) is the radiative (non-radiative) decay rate and the lifetime of the excited state is given by $\tau_{\rm{fs}} = \frac{1}{\gamma_{\rm{fs}}} = \frac{1}{\gamma_{\rm{r}} + \gamma_{\rm{nr}}}$. As the Purcell effect only affects the radiative decay channel, the total decay rate in the presence of a dielectric environment which changes the optical density of states is given by $\gamma = F'_{\rm{theo}} \gamma_{\rm{r}} + \gamma_{\rm{nr}}$ (assuming that the non-radiative rate is unchanged by the presence of the dielectrics and optimal emitter-mode coupling, ie. $\upsilon = \xi = \chi = 1$). In particular the choice $\xi = 1$ implies that the transition dipole of the emitter is aligned parallel to the planar mirror surface of the cavity ($\gamma_p$ in Fig. \ref{figS3}).
We can thus write down the decay rates of an emitter positioned at the mirror surface radiating into the half-space above $\gamma_{\rm{hs}}$ and an emitter within the cavity $\gamma_{\rm{cav}}$:
\begin{align}
\gamma_{\rm{hs}} &= \mathcal{E} \gamma_{\rm{r}} + \gamma_{\rm{nr}} \label{eqRates1}\\
\gamma_{\rm{cav}} &= F'_{\rm{theo}} \gamma_{\rm{r}} + \gamma_{\rm{nr}}
\label{eqRates2}
\end{align}
Here $\mathcal{E}$ quantifies the enhancement (and suppression) of the radiative decay rate for a dipole positioned parallel to the surface of the low terminated DBR.

The effective Purcell factor measured in the experiment is given by $F'_{\rm{exp}} = \gamma_{\rm{cav}}/\gamma_{\rm{fs}} = F_{\rm{exp}} + 1$. Experimentally we obtain the ratio between the two rates given in Eq. \ref{eqRates1} and \ref{eqRates2}, ie. 
\begin{equation}
f = \frac{\gamma_{\rm{cav}}}{\gamma_{\rm{hs}}} = F'_{\rm{exp}} \frac{\gamma_{\rm{fs}}}{\gamma_{\rm{hs}}} =   \frac{(F'_{\rm{theo}} - 1)\eta + 1}{(\mathcal{E} - 1)\eta + 1}
\end{equation} 
where we have used the definition of the quantum efficiency $\eta$. We call $f$ the measured enhancement. $F'_{\rm{exp}}$ is thus given as $F'_{\rm{exp}} = f(\mathcal{E}-1)\eta +f$. For the quantum efficiency $\eta$ we obtain:
\begin{equation}
\eta = \frac{F_{\rm{exp}}}{F_{\rm{theo}}}  = \frac{f -1}{ f -1 + F'_{\rm{theo}} - \mathcal{E} f}
\label{eta}
\end{equation} 

\section{Collection efficiency}
The collection efficiency is given by $\eta_{\rm{fs}} = 2 \frac{\Omega_{\rm{NA}}}{4\pi} =  1 - \sqrt{1-(\rm{NA}/n)^2}$, where $\Omega_{\rm{NA}}$ is the solid angle subtended by the lens, $n$ is the refractive index of the medium between mirror and lens and the factor $2$ arises from the reflection off the mirror. The divergence of the cavity modes in this study is below $\theta_{\rm{div}} = 0.4$, which suggests that all the light is collected from the cavity mode irradiating through the planar mirror side. Since the reflectivity of both DBRs is the same, we obtain a collection efficiency of $\eta_{\rm{cav}} = 50 \%$ for emission out of the cavity.

\section{Second-order degree of coherence}
We employ a Hanbury Brown and Twiss setup to obtain a histogram of the times between detection events of two SPADs (Perkin Elmer SPCM-AQRH). The detectors are connected to the two ends of a fiber splitter which receives photons from the cavity setup through the collection optics of the confocal microscope. Depending on the width $\Delta t$ of the timing bins chosen to build up the histogram such measurements can take a long time: The coincidence countrate $\Gamma_{\rm{CCR}}$ per bin is given as
\begin{align}
\Gamma_{\rm{CCR}} = \Gamma_{\rm{CR1}} \Gamma_{\rm{CR2}} \Delta t
\end{align}
Here $\Gamma_{\rm{CR1}}$, $\Gamma_{\rm{CR2}}$ are the average countrates on detector 1 and 2. The experimental data can be normalised to obtain the second-order degree of coherence $g^{(2)}(\tau)$ by dividing the counts in each bin by $N = \Gamma_{\rm{CCR}} T$, where $T$ is the total acquisition time. For the data presented in this work an equation derived with a three level model is used to fit the experimental data for continuous wave excitation:
\begin{align}
g^{(2)}(\tau) = 1 + p\left( b e^{-|\tau|/\tau_2} -(1+b)e^{-|\tau|/\tau_1} \right)
\end{align}
Here $p$ is the probability that the photon was emitted by the single emitter ($p > 0.5$ implies $g^{(2)}(0) < 0.5$), $b$ describes the bunching around the central dip typical to a three-level system at large pumping powers and $\tau_1$, $\tau_2$ are the associated time scales of the radiative decay and the bunching processes \cite{kurtsiefer_stable_2000,kaupp_purcell-enhanced_2016}.

For the data presented in Fig. 3a of the main text the experimental parameters were $\Gamma_{\rm{CR1}} = 12.92 \frac{\rm{kcts}}{\SI{}{\second}}$, $\Gamma_{\rm{CR2}} = 12.78 \frac{\rm{kcts}}{\SI{}{\second}}$, $\Delta t = \SI{0.244}{\nano\second}$ and $T = \SI{50}{\min}$. The fit parameters were obtained as $b = (0.15 \pm 0.01)$, $p = (0.70 \pm 0.09)$, $\tau_1 = (1.35 \pm 0.23)$ $\SI{}{\nano\second}$ and $\tau_2 = (75 \pm 22)$ $\SI{}{\nano\second}$ and a fit range of \SI{200}{\nano\second} around $\tau = \SI{0}{\nano\second}$ was used. 

\putbib
 
\end{bibunit}

\end{document}